# Investigation of enhanced second harmonic generation in laser-induced air plasma

SHING YIU FU,[1] KAREEM J. GARRIGA FRANCIS,[1] MERVIN LIM PAC CHONG,[2] YIWEN E,[1] X.-C. ZHANG[1,*]

[1]*The Institute of Optics, University of Rochester, New York 14627, USA*
[2]*Laboratory for Laser Energetics, University of Rochester, Rochester, NY 14623, USA*
*Corresponding author: zhangxc@rochester.edu*

**We report a systematic investigation into the processes behind a near hundredfold enhanced second harmonic wave generated from a laser-induced air plasma, by examining the temporal dynamics of the frequency conversion processes, and the polarization of the emitted second harmonic beam. Contrary to typical nonlinear optical processes, the enhanced second harmonic generation efficiency is only observed within a sub-picosecond time window and found to be nearly constant across fundamental pulse durations spanning from 0.1 ps to over 2 ps. We further demonstrate that with the adopted orthogonal pump-probe configuration, the polarization of second harmonic field exhibits a complex dependence on the polarization of both input fundamental beams, contrasting with most of the previous experiments with a single-beam geometry.**

Since the advent of laser more than half a century ago, extensive research has been carried out to develop the theory behind second harmonic generation (SHG), and to exploit its far-reaching potential. SHG has found numerous applications in laser technology advancement, optical communications and sample characterization. In particular, with the ripening of table-top lasers operating at high power and a catalog of wavelengths, the development of robust and nondestructive SHG methods is enabled for assaying a span of optical materials, biological specimens, and even plasma systems [1-6]. SHG and its polarization serve as distinctive indicators in characterizing the nonlinear properties of a target such as optical anisotropy [1-4], structural and composition variants of molecules [5,7,8] and plasma electron density fluctuations [6].

For an isotropic medium like atomic vapor or gas, conventional wisdom may demand the second-order nonlinear susceptibility, $\chi^{(2)}$, of the material to vanish under the dipole approximation, due to its microscopic centrosymmetric atomic arrangement [9]. However, when a high-peak-power laser is used to ionize the media and create a plasma, SHG at unanticipated efficiencies has been observed, seemingly violating the prediction following the vanishing $\chi^{(2)}$ noted above. Current explanations for such SHG in femtosecond (fs) laser-induced plasma experiments include air lasing [10], electric field-induced second harmonic generation (EFISH) effects which leverage the material $\chi^{(3)}$ [11-13], and inhomogeneous plasma $\chi^{(2)}$ arising from the broken symmetry and plasma gradient [14-16]. It has also been demonstrated that if the contribution of quadrupole transition is included in addition to the dipole approximation, one can account for second-order optical nonlinear effects in an isotropic atomic vapor [17].

In a recent report, a remarkably efficient SHG process in a laser-induced air plasma showcased a 0.02% power efficiency [13]. Contrasting with most of the previous studies in gaseous media [10,11,14-16,18-20], a pump-probe configuration is adopted, producing a strong SH signal that is temporally constrained by the pump-probe interaction window. The SH signal is thus only detected over a sub-picosecond (ps) timespan, which is much shorter than the expected plasma relaxation time on the nanosecond scale [21].

This letter seeks to delineate the SHG properties under such orthogonal pump-probe excitation scheme. By mixing an incrementally time-delayed optical probe beam into a laser-induced, under-dense air plasma, we experimentally analyze the temporal dynamics and polarization of the generated SH beam. With sufficiently long optical pulses, the SHG can be differentiated from concurrent frequency conversion processes. These identified SH signals are compared against results presented and predicted in previous studies. Curiously, the efficiency of the SHG process of interest is chiefly insensitive to pulse duration over a multi-ps range, and is boosted by almost 100 times in the presence of the air plasma. Furthermore, while the SH polarization exhibits simultaneous dependences on both the pump and probe polarization angles, it in general possesses both P and S components. The polarization is thus not co-polarized with either linearly polarized fundamental optical excitation field, setting the process apart from most of the past investigations.

The experimental setup is schematically illustrated in **Fig. 1(a)**. To facilitate an efficient and readily repeatable investigation, a single-color air plasma source is adopted. The plasma-inducing pump beam and SH-generating probe beam are provided by a Coherent Astrella system, which outputs 800nm laser pulses at 1 kHz repetition rate with variable pulse duration. The pump pulse carries 1.4 mJ pulse energy and creates a visible filament as captured in **Fig. 1(a).** The split-off probe pulse possesses an attenuated 0.3 mJ pulse energy, in pursuit of suppressing the background noise along z-direction in the forms of plasma fluorescence and supercontinuum generation. A downstream analyzer unit further isolates the SH beam for characterization with a photomultiplier tube (PMT). To investigate the SHG dynamics over a moving temporal overlap between the pump and probe pulses, a motorized linear delay stage is available to control the time delay (τ) between the two paths. The inset of **Fig. 1(a)** shows that when the two pulses temporally and spatially overlap, a drastically increased fluorescence is observed. The bright dot beneath the obvious filament highlights this interaction volume. Finally, to enable the study on SHG sensitivity to fundamental beam polarization, two half-wave plates (HWP) are inserted to independently rotate the pump (γ) and probe polarization (φ) angle respectively, during a measurement as depicted in **Fig. 1(b)**.

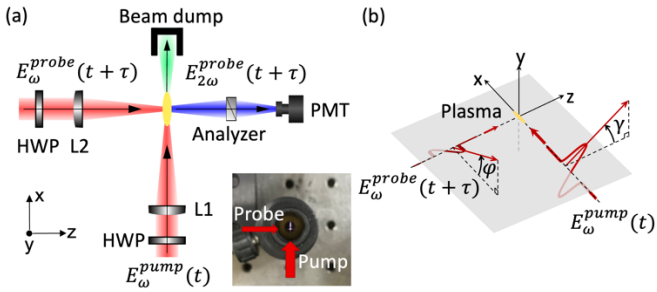

**Fig. 1.** (a) Experimental setup of the pump-probe SHG in laser-induced air plasma. The pump ($E_\omega^{pump}$) and probe ($E_\omega^{probe}$) pulses are offset by a time delay, τ. The inset shows a photograph of the beam intersection region, displaying the pump laser filament and the dot of pump-probe interaction volume underneath. HWP: half-wave plate. L1: 125-mm lens. L2: 150-mm lens. Analyzer: 400 nm bandpass filters, a Glan prism polarizer and a focusing lens. (b) Definition of the pump (γ) and probe beam (φ) polarizations respectively, following the coordinate system in (a).

We first characterize the evolution of the SH signal as the temporal overlap between the pulse shifts. **Fig. 2(a)** shows the SH emission intensity versus the time delay, τ, between the two beams. Different colors indicate the results of varied pulse duration by adding chirp to the pulses. A scrutiny of the data with pulse duration over 2 ps reveals an apparent decomposition of signal into two regimes: a sharply spiked enhancement spanning a sub-ps time window near τ = 0 and a broad baseline emission whose width closely correlates with the fundamental pulse duration. Interestingly, both timescales are orders-of-magnitude shorter than the nanosecond plasma lifetime. In particular, the enhancement spike appears to further slim down over τ as the pulse duration of the fundamental optical beam increases. The narrow interaction windows suggest that neither regime shall simply stem from a straightforward $\chi^{(2)}$ process in the slowly decaying plasma.

Another distinguishing feature resides in the peak intensity of the enhanced spikes, projected on and traced by the red dashed line. The SH intensity remains largely constant for pulses from 0.1 ps to over 2 ps, and is only halved when excited with 6-ps pulses. This trend is an alerting distinguishment from typical nonlinear processes, for instance, in an SHG autocorrelation measurement that quantifies ultrafast pulse durations, as shown in **Fig. 2(b)**. In the presented measurement, two collinear 800 nm optical beams with a variable time delay interact inside a beta barium borate crystal (β-BBO), resulting in a boosted SH signal as the time delay between the two beams approaches zero. It is obvious that the signal strength, and the underlying nonlinear interaction efficiency, would decline much more sharply with stretched pulses, due to the power law between nonlinear polarization and peak fundamental electric field. Hence, comparing the peak intensity projection versus pulse duration in both cases in **Fig. 2**, we can see that the SHG from air plasma abnormally sustains its intensity over a much longer pulse before dipping.

It is also imperative to recall the propagation direction of the SH beam. In a conventional crossed-beam autocorrelation measurement, the time delay-sensitive SHG favors an emission along the bisector of the two k-vectors of the fundamental beams, due to phase-matching considerations. In contrast, in **Fig. 2(a)** and the previous report [13], the unexpectedly strong SH component is measured along the probe path and modulated within a sub-ps τ-window. This further hints at a difference in generation processes behind the signals in **Fig. 2(a)** and **(b)**, and potentially an ultrafast plasma property to be identified during the SHG enhancement process.

To further dissect the sub-ps temporal dynamics of the SHG enhancement, we examine the SH signal generated with 4.1 ps fundamental pulse duration in **Fig. 3(a)**. The plot identifies three SHG windows. As the optical pump pulse much precedes the probe pulse and there is negligible temporal overlap in Section III, the noise-like signal is the SHG solely from a focusing probe beam, similar to past literature with a single-beam configuration [11,12,14-16,18]. Section II is the plasma fluorescence that gradually grows as the two pulses temporally superpose. Section I contains the SHG enhancement of interest, which exists only within a sub-ps window. The contrast in signal strengths between Section I and III underscores the SHG enhancement by the presence of a laser-induced air plasma. A separate measurement by blocking the pump beam records an 80-time difference in SHG strength between the two cases.

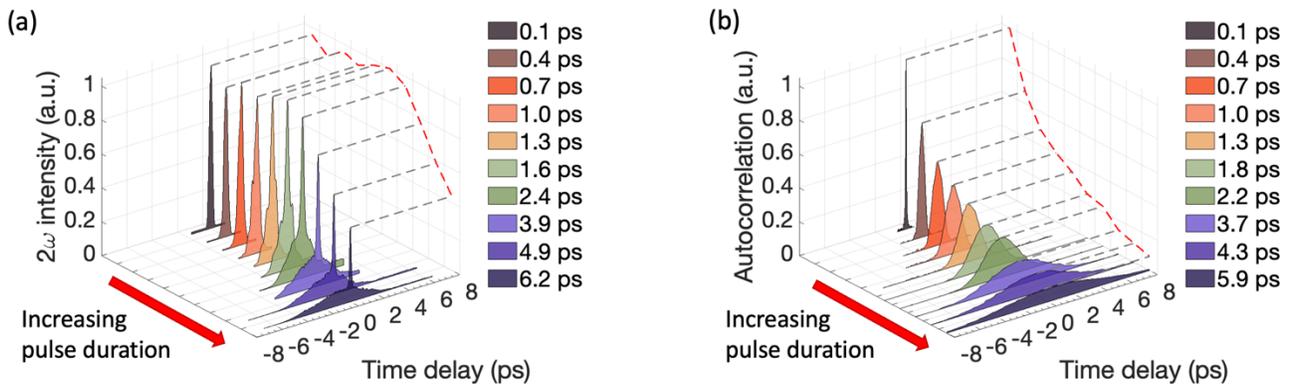

**Fig. 2.** (a) SH signals measured with the orthogonal pump-probe setup with various fundamental pulse durations. (b) A separate measurement of SHG autocorrelation signals that reflect the fundamental pulse duration. The peak values of the signals are projected onto the red dashed lines to highlight the distinctive trends. In both sub-figures, the pulses in both pump and probe beams are chirped simultaneously, thus sharing the same pulse duration as indicated by the legend entries.

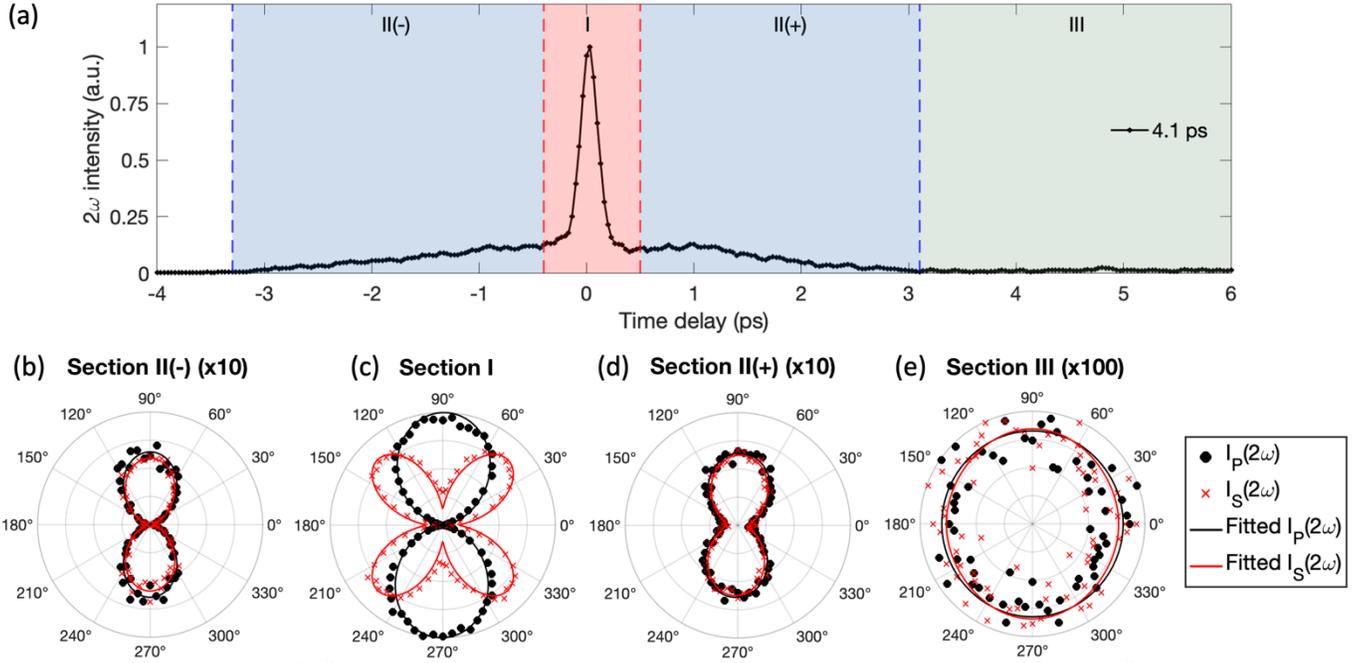

**Fig. 3.** (a) Temporal dynamics of SHG with 4.1 ps optical pulses. The signal is partitioned into sections I (Red), II (Blue) and III (Green). (b-e) Transition of SH polarization with respect to fundamental beam polarizations. The polar coordinate represents a full rotation of the probe polarization, φ. Pump polarization, γ, is fixed at 90°. P-polarized 2ω intensity component (Black dots) is defined to be parallel to the x-z plane, while S component (Red crosses) points along y-axis. To find a working functional form of the polarization dependence, the data are fitted with sinusoidal functions plotted in solid lines (Equations S1 and S2). (b) SH Polarization in Section II(-), (c) in Section I, (d) in Section II(+), (e) in Section III. Intensities in Section II and III are magnified 10 and 100 times respectively for clarity.

To provide an alternative perspective for analyzing the difference between the three sections, we carry out a polarization study on the SH beam. **Fig. 3(b)** to **(e)** highlights the dependence of SH polarization on the input polarization combinations in the three windows respectively. The SH field polarization is decomposed into two orthogonal directions, P and S, as defined in **Fig. 3** captions. In **Fig. 3(b) and (d)**, it can be observed that both SH components share a preference for S-polarized probe pulse, which is when the plasma fluorescence maximizes at the interaction volume. In Section III, on the other hand, the self SHG of probe beam has insufficient signal-to-noise ratio to conclude its polarization state, and appears largely unpolarized at all φ. **Fig. 3(e)**, however, uncovers a four-leaved dependence of the S component on φ, on top of a ten to hundredfold intensity amplification as compared to Section II and III. This discovery strengthens the hypothesis that the SHG enhancement in Section I originates from a distinct nonlinear process yet to be identified. This four-lobe pattern is invariant over the entire sub-ps time window or as the plasma position shifts relative to the intersection spot with the probe beam (see **Fig. S2**).

A more deliberate polarization study on Section I SHG is performed in **Fig. 4**. Across the nine sub-figures, the pump polarization angle, γ, is incremented in steps of 20°. The probe polarization, φ, is again indicated by the polar coordinate. It is immediately apparent that the intensities of both P and S components of the SH wave are minimal when γ = 0°. With a little more care in interpreting the subsequent data sets, we can also see significantly reduced signal strengths whenever φ = 0 or 180°, which are exactly when the polarization direction of probe beam is parallel to the propagation direction of the pump. It is hence hypothesized that since a propagating wave does not support a polarization component parallel to its propagation direction, the incompatibility first reduces the interaction strength between the two beams, and ultimately the nonlinear process efficiency.

A more peculiar phenomenon lies in the shifting polar dependence of the intensities. If γ is fixed at 20°, the four-lobe curve seen in **Fig. 3(c)** is identified in the P component of the SH beam instead. There are four angles of φ at which the P intensity reaches local maximum. However, as γ is incremented, the four-lobe shape gradually reduces into a figure-eight. There are only two angles of φ at which the P component reaches maximum when γ is near 90°, or S-polarized. As γ is further increased, the two lobes revert to four lobes. On the contrary, the S component starts off with two lobes when γ is near integer multiples of π, and shifts into four lobes.

Perhaps surprisingly, the anti-coupled oscillation between a four-lobe and two-lobe dependences in the P and S components is also observable after a swap between the pump and probe (see **Fig. S1**). When the probe polarization is fixed and progressively incremented, the same alternations between two-lobe and four-lobe polar dependences on the pump polarization, γ, can be detected. This behavior, while puzzling to be observed from the SHG in an air plasma source, is well-established in SHG characterization of crystal solids [1-4]. A setup with a piece of nonlinear crystal replacing the pump plasma can display highly resembling plots when the crystal orientation is rotated in steps [4], similar to pump beam polarization incrementation in **Fig. 4**. Nevertheless, provided that a microscopic crystalline structure in air plasma is improbable, this resemblance warrants further investigation into the SHG within this narrow section.

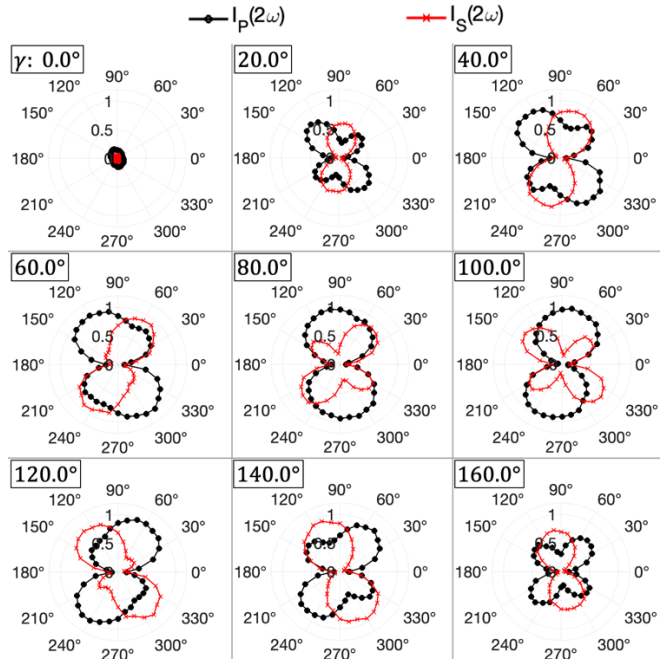

**Fig. 4**. Evolution of the P and S SH polarization component intensities with respect to pump polarization, γ, and probe polarization, φ. γ is incremented from 0 to 160° in steps of 20°. The polar coordinate corresponds to a full 360° rotation on φ.

In this study, we report the time delay dynamics and polarization of the SH wave generated inside a laser-induced air plasma utilizing a pump-probe setup. The directional SH beam follows the probe path with a near hundredfold enhancement observed within a sub-ps pump-probe overlap window. The short-lived process and its strength show a surprising resistance to varying laser pulse durations and can be picked out among other concurrent SHG processes, in the forms of plasma fluorescence and the background self SHG by the focused probe optical beam. The enhanced SH signal possesses polarization properties that contrast with those from simultaneous SHG processes and past investigations on SHG in gaseous media. The polarization has been demonstrated to depend heavily on both pump and probe fundamental optical beam polarizations. Remarkably, with a rotation of either fundamental beam polarization, the polarization measurement has revealed a surprising resemblance to reported data in similar experiments where a rotating solid crystal sample is in place of the plasma. The narrow generation window, sustained interaction strength at long pulse durations and polarization properties of the enhanced SH beam all call for further investigation into this exciting phenomenon.

**Funding.** Air Force Office of Scientific Research (FA9550-21-1-0389); National Science Foundation (ECCS-2152081).

**Disclosures.** The authors declare no conflicts of interest.

**Data availability.** Data underlying the results presented in this paper are not publicly available at this time but may be obtained from the authors upon reasonable request.

**Supplemental document.** See Supplement 1 for supporting content.

# Investigation of enhanced second harmonic generation in laser-induced air plasma: supplemental document

## 1 FITTING EQUATIONS

Least squares fitting of sinusoids for two and four-lobe dependences in **Fig. 3**, where $k_i$ are fitting parameters.

$$I_{4-lobe} = |k_1 \sin(2\varphi + k_2 \sin(2\varphi))|^{k_3} + k_4. \quad (S1)$$
$$I_{2-lobe} = |k_1 \sin(\varphi + k_2)|^{k_3} + k_4. \quad (S2)$$

## 2 EVOLUTION OF SECOND HARMONIC FIELD POLARIZATION WITH INCREMENTAL PROBE BEAM POLARIZATION

Anti-coupled oscillation of P and S components of the second harmonic emission between a two-lobe and four-lobe dependences on pump polarization angle (γ), as probe polarization angle (φ) is fixed and incremented from 0° to 160°. This information is in fact already embedded and hidden in **Fig. 4** of the main text, but is replotted here with a swap of roles between γ and φ for a more intuitive inspection.

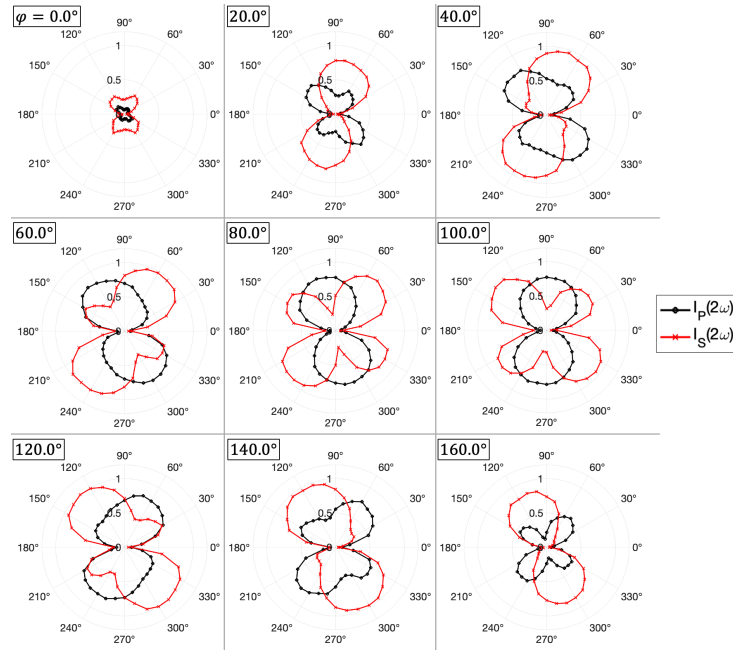

**Fig. S1**. Evolution of the P and S second harmonic polarization component intensities with respect to pump polarization, γ, and probe polarization, φ. φ is incremented from 0 to 160° in steps of 20°. The polar coordinate corresponds to a full 360° rotation on γ.

# 3 SHG WITH SCANNING TIME DELAY AND PUMP PLASMA POSITION

**Fig. S2** shows that the multi-lobe patterns in **Fig. 3** are angle-invariant with respect to time delay, $\tau$, and pump plasma position along the x-axis with near 100 fs pump-probe pulse duration. The pump plasma can be translated with the focusing lens L1 in **Fig. 1**. For simplicity, only the S component of the second harmonic wave is plotted, as the P component also follows the identical trend. When either the time delay or pump plasma location is shifted from the optimum location, the temporal or spatial overlaps of the optical pulses deteriorate, accounting for a narrowed interaction window and consequently the decline in measured power. The probe polarization $\varphi$-invariance reassures that the generation mechanism has not changed during this window of concern.

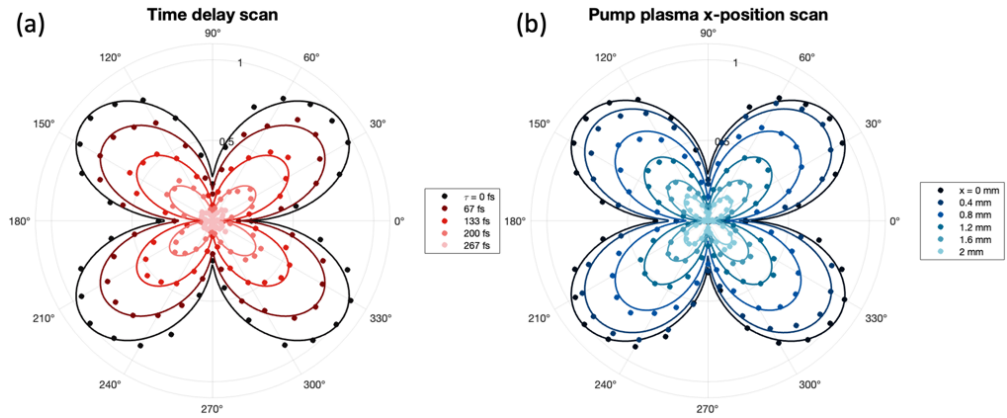

**Fig. S2**. The dependences of SH S component intensity on probe polarization, $\varphi$, with pump polarization, $\gamma$, fixed at 90° (a) at various pump-probe time delays, $\tau$, and (b) at various pump plasma translations along x. The dots represent the experimental data points, which are fitted to sinusoidal functions (Equation S1 and S2) traced by the solid lines.